\documentclass[11pt,twoside,a4paper,cr,final]{cms-tdr}
\def\svnVersion{66787MP}

\begin{document}

\hyphenation{had-ron-i-za-tion}
\hyphenation{cal-or-i-me-ter}
\hyphenation{de-vices}

\RCS$Revision: 62698 $
\RCS$HeadURL: svn+ssh://svn.cern.ch/reps/tdr2/papers/XXX-08-000/trunk/XXX-08-000.tex $
\RCS$Id: XXX-08-000.tex 62698 2011-06-21 00:28:58Z alverson $
\cmsNoteHeader{CR-2011/097} 
\title{Centrality and $\PT$ dependence of charged particle $\RAA$ in PbPb collisions
at $\sqrt{s_{_{\mathrm{NN}}}}$ = 2.76 TeV}

\author[mit]{Andre S. Yoon for the CMS collaboration}

\date{\today}

\abstract{
The transverse momentum ($p_T$) spectra of charged particles is measured by CMS as a function of collision centrality in PbPb
collisions at $\sqrt{s_{_{NN}}} = 2.76$ TeV. The results are compared to a pp reference spectrum, constructed by interpolation
between $\sqrt{s}$ = 0.9 and 7 TeV measurements. The nuclear modification factor ($R_{AA}$) is constructed by dividing the
PbPb $p_T$ spectrum, normalized to the number of binary collisions ($N_{coll}$), by the pp spectrum. Measured $R_{AA}$ in 0--5\%
centrality bin is compared to several theoretical predictions.

\vspace{8mm}

\begin{center}
Presented at \textit{QM2011}: \textit{Quark Matter 2011}
\end{center}
}

\newcommand {\XT}          {\ensuremath{x_T}}
\newcommand{\RAA}        {\ensuremath{R_{AA}}}
\newcommand{\TAA}        {\ensuremath{T_{AA}}}
\newcommand{\Npart}      {\ensuremath{N_{part}}}
\newcommand{\Ncoll}       {\ensuremath{N_{coll}}}

\newcommand{\microbinv} {\mbox{\ensuremath{\,\mu\text{b}^\text{$-$1}}}\xspace}

\hypersetup{%
}

\maketitle 

\section{Introduction}
The charged particle $\PT$ spectrum is an important observable for studying the properties of the hot, dense medium 
produced in the collisions of heavy nuclei. In particular, the modification of the $\PT$ spectrum compared to nucleon-nucleon collisions 
at the same energy can shed light on the detailed mechanism by which hard partons lose energy traversing the medium~\cite{dEnterria:2009am}.
The modification is typically expressed in terms of 
\begin{equation}
\RAA(\PT) = \frac{d^{2}N_{\mathrm{ch}}^{\mathrm{AA}}/d\PT d\eta}{\left<\TAA\right> d^{2}\sigma_{\mathrm{ch}}^{\mathrm{NN}}/d\PT d\eta},
\label{eqn:def_raa}
\end{equation}

where $N_{\mathrm{ch}}^{\mathrm{AA}}$ and $\sigma_{\mathrm{ch}}^{\mathrm{NN}}$ represent the charged particle yield in 
nucleus-nucleus collisions and the cross section in nucleon-nucleon collisions, respectively. The nuclear overlap function
$\left<\TAA\right>$ is the ratio of the number of binary nucleon-nucleon collisions $\left<\Ncoll\right>$ calculated from a Glauber model 
of the nuclear collision geometry. At RHIC, the factor of five suppression seen in $\RAA$ was an early indication of strong final-state medium effects 
on particle production. In this contribution, measured charged particle spectra and $\RAA$ in bins of centrality 
for the $\PT$ range of 1--100 $\GeVc$~\cite{CMARaa} are presented.

\section{Experimental Methods}
This measurement is based on a data sample corresponding to an integrated luminosity of 7 $\microbinv$ collected by the CMS experiment~\cite{JINST} in 2.76 TeV PbPb collisions. 
A minimum bias event sample was collected based on a trigger requiring a coincidence between signals in the +$z$ and \mbox{--$z$} sides of either the forward hadronic calorimeters (HF) or the beam scintillator counters (BSC). To ensure a pure sample of inelastic hadronic collision events, additional offline selections were performed. These include a beam halo veto based on the BSC timing, an offline HF coincidence of at least three towers ($E > 3$ GeV) on each side of the interaction point, a reconstructed vertex consisting of at least two pixel tracks of $\PT > 75$ $\MeVc$, and a rejection of beam-scraping events based on the compatibility of pixel cluster shapes with the reconstructed primary vertex. The collision event centrality is determined on the total energy deposited in both HF calorimeters. 

In this analysis, data recorded by single-jet triggers with uncorrected transverse energy thresholds of $E_{T} = 35$ GeV (Jet35U) and 50 GeV (Jet50U) are included in order to extend the statistical reach of the $\PT$ spectra. The use of the jet-triggered data, which contributes dominantly to the high-$\PT$ region ($\PT > 50$ $\GeVc$), also allows to keep low misidentification rates of the selected tracks being enforced by an implicit track-calorimeter matching. The jet-energy thresholds in the trigger are applied after subtracting the underlying event energy but without correcting for calorimeter response. $E_{T}$ distributions of the most energetic reconstructed jet with $|\eta|<2$, referred to as the leading jet, are normalized per minimum bias event. Following the procedure introduced in the analogous measurement of the charged particle spectra in 0.9 and 7 TeV pp collisions~\cite{Chatrchyan:2011av}, the spectra are calculated separately in three ranges of leading-jet $E_T$, each corresponding to a fully efficient trigger path, and then combined to obtain the final result. 

\begin{figure}[ht!]
\begin{center}
   \includegraphics[width=0.46\textwidth]{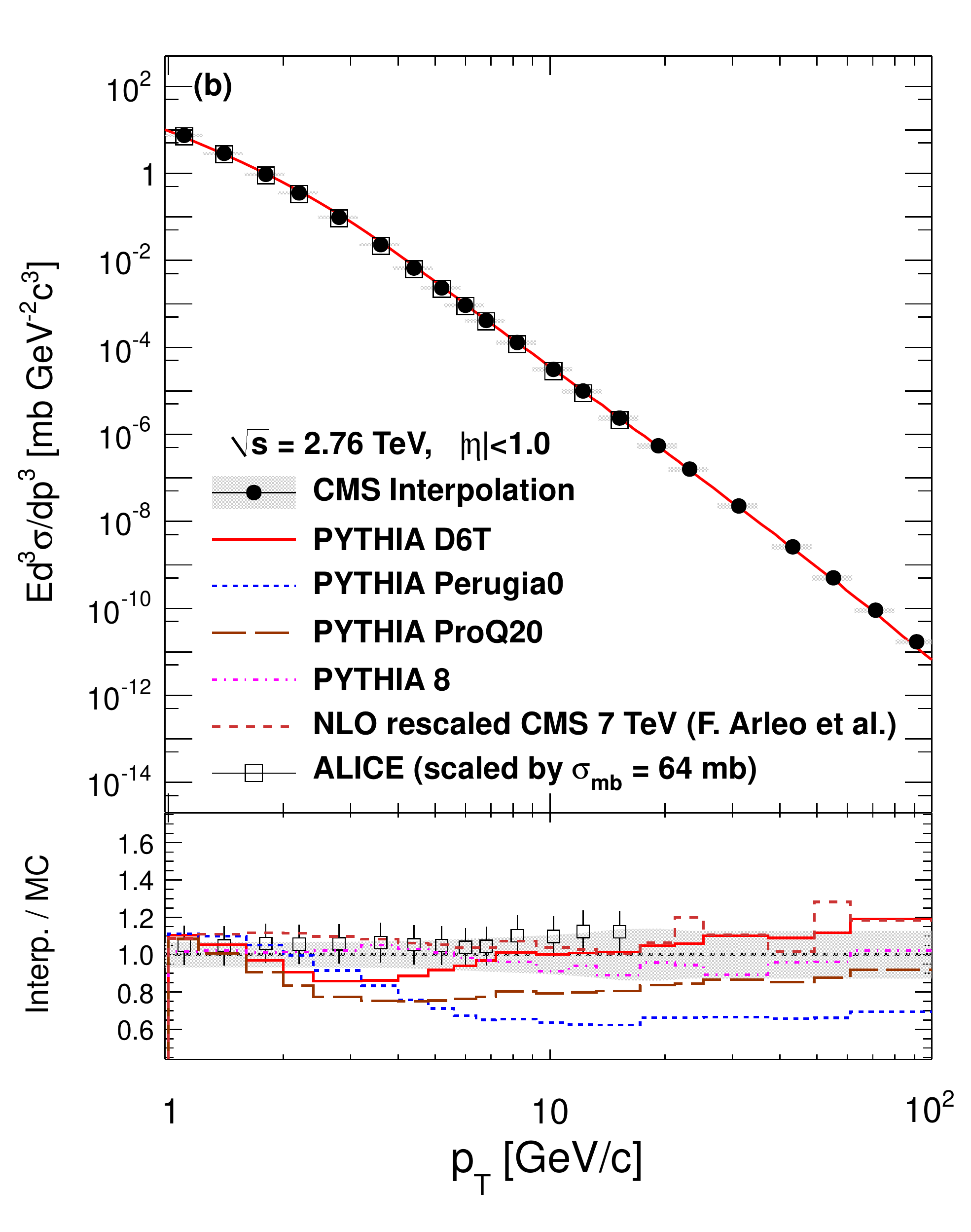}
       \hspace{6mm}
       \includegraphics[width=0.46\textwidth]{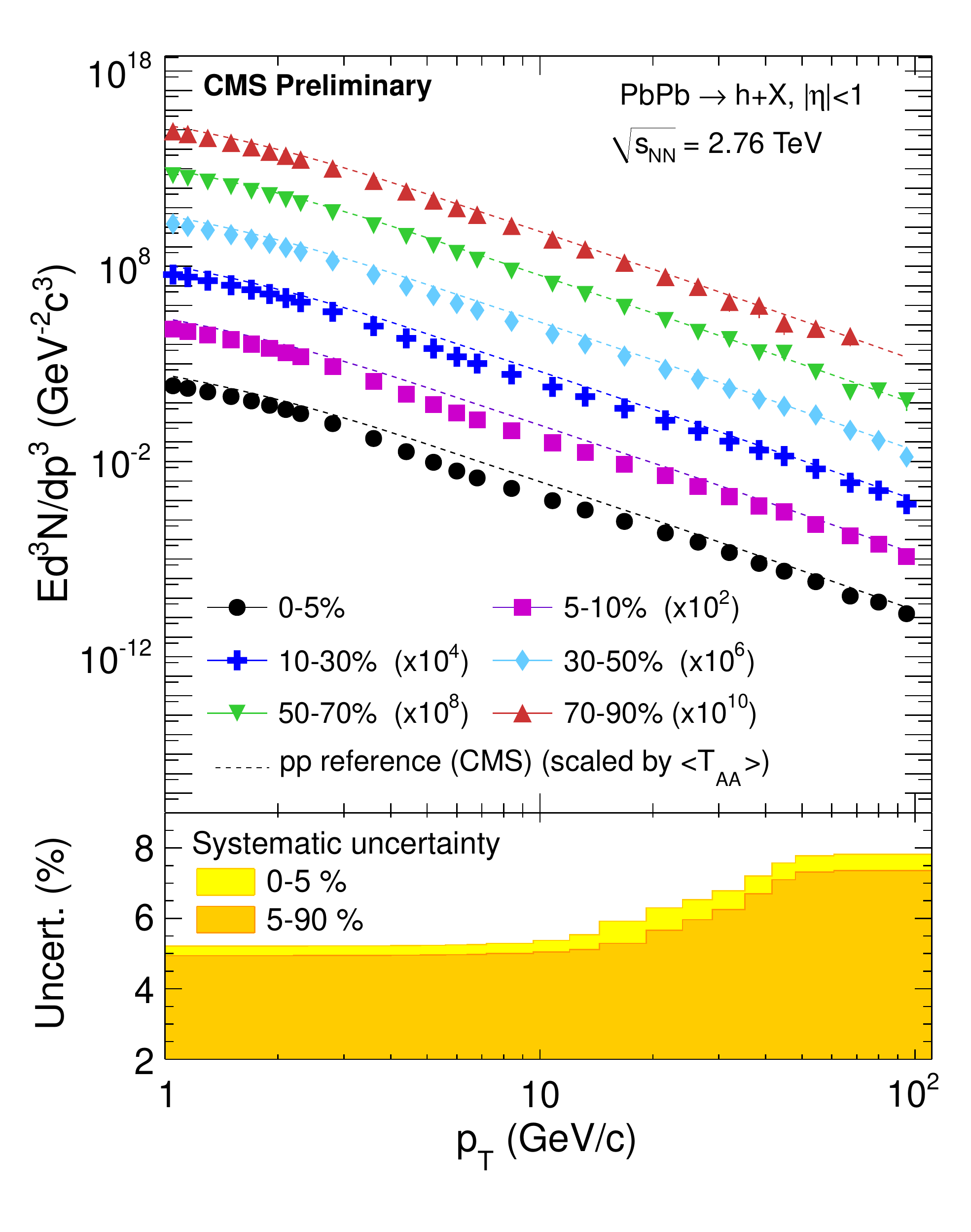} 
       \caption{ 
       (Left) Interpolated 2.76 TeV charged particle differential transverse momentum cross section with ratios of combined interpolation to various predictions and interpolation. 
       (Right) Invariant differential yield in bins of collision event centrality (symbols), compared to a pp reference spectrum, scaled by the corresponding number of binary nucleon-nucleon collisions (dashed lines). 
The systematic uncertainties on the PbPb differential yields as a function of $\PT$ for the 0--5\% and 5--90\% are shown in the bottom panel.}
\label{fig:specInvLogX}
\end{center}
\end{figure}

\section{Results}
The interpolated differential cross section in pp at 2.76 TeV~\cite{Chatrchyan:2011av} is shown in the upper panel of Fig.~\ref{fig:specInvLogX}, and its ratio with respect to various \textsc{pythia} tunes.
Also shown in the lower panel of Fig.~\ref{fig:specInvLogX} is the ratio of the predicted 2.76 TeV cross section to that found by
simply scaling the CMS measured 7 TeV result by the expected 2.75 TeV to 7 TeV ratio from NLO calculations~\cite{Arleo:2010kw} and the interpolation used in the recent \mbox{ALICE} publication~\cite{aliceRAA}. 

The charged particle invariant differential yields are shown for the six centrality bins in Fig.~\ref{fig:specInvLogX} and compared
to the corresponding quantity taken from the interpolated pp reference spectrum~\cite{Chatrchyan:2011av}, normalized by the number of binary collisions. The points have been scaled by the factors given in the figure for easier viewing. 

\begin{figure}[ht!]
   \begin{center}
        \includegraphics[width=0.59\textwidth]{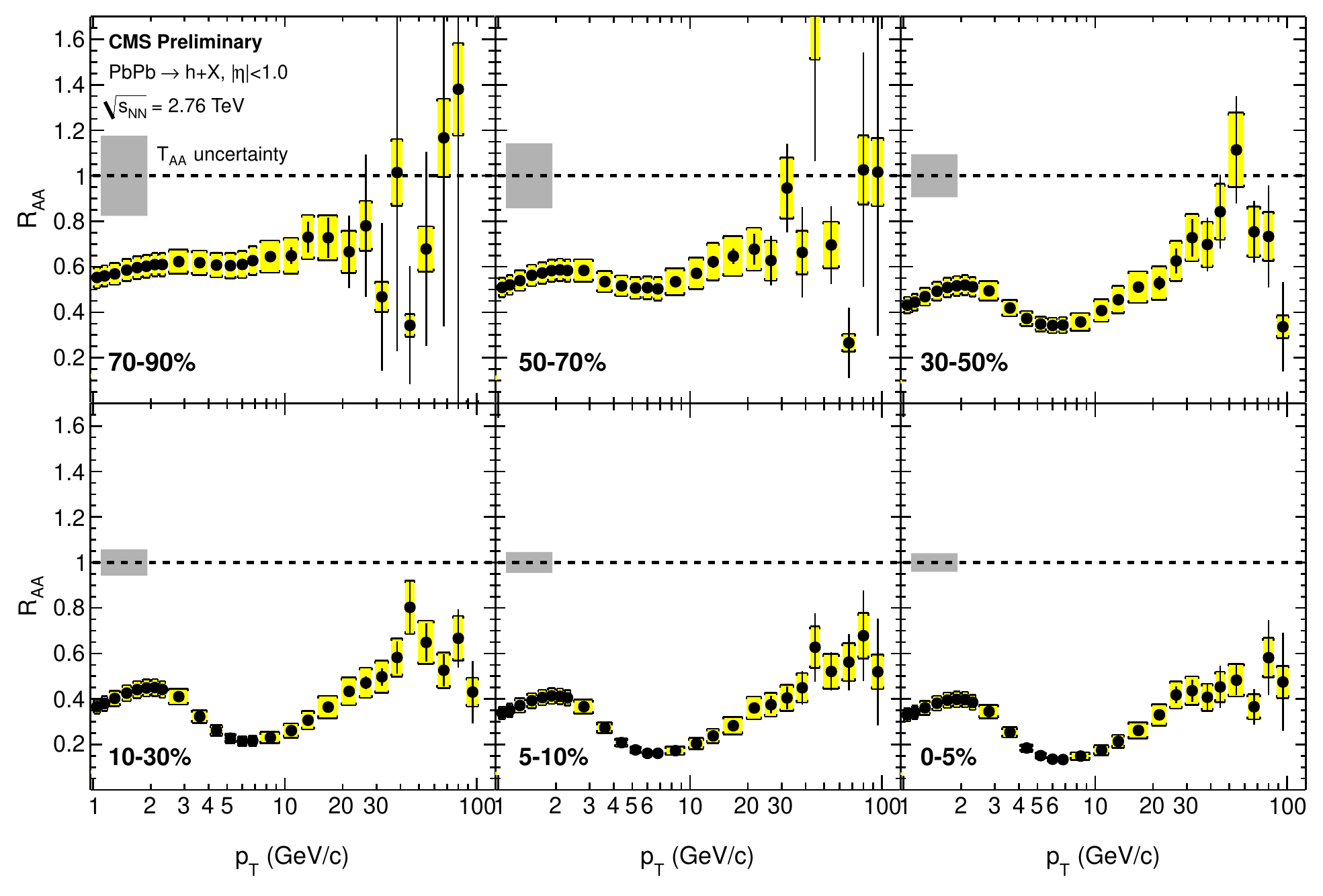} 
        \includegraphics[width=0.402\textwidth]{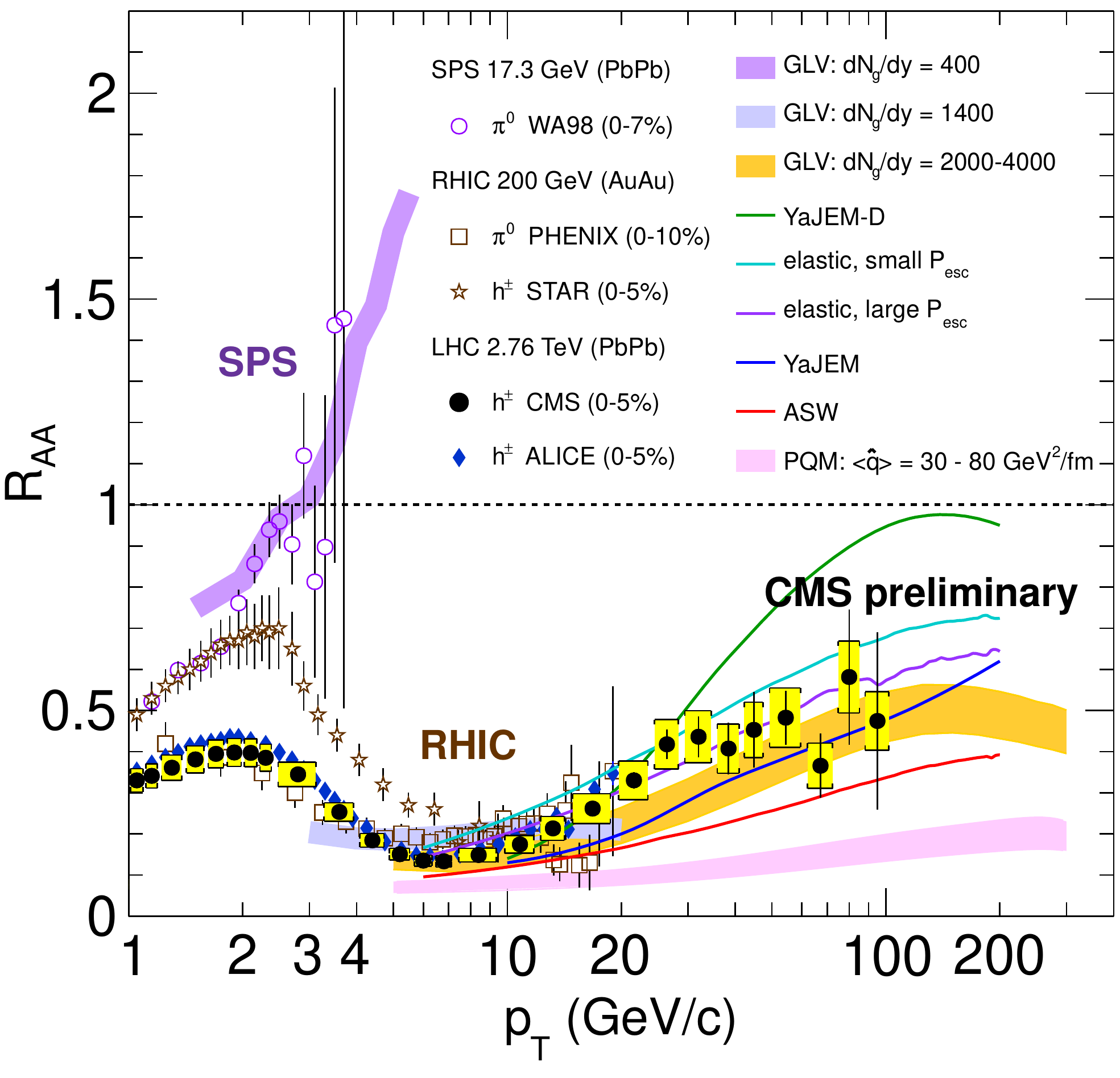}
	\caption{(Left) $\RAA$ (filled circles) as a function of $\PT$ for six centrality intervals. The error bars represent the statistical uncertainties, and the yellow boxes the $\PT$-dependent systematic uncertainties on the $\RAA$ measurements.  
      (Right) Comparison to the measurements of $\RAA$ in central heavy ion collisions at three different center-of-mass energies  as a function of $\PT$ for neutral pions and charged hadrons and to several theoretical predictions~\cite{pqm,glv,renk}.}
     \label{fig:specRaaLogX}
   \end{center}
\end{figure}

$\RAA$ is presented as a function of transverse momentum in Fig.~\ref{fig:specRaaLogX} for each of the six centrality bins.
The shaded areas (boxes) around the points show the systematic uncertainties including those from the interpolated pp reference spectrum.  An additional systematic uncertainty from the $\TAA$ normalization, common to all points is displayed as the shaded band around unity in each plot. In the most-peripheral events (70--90\%), a moderate suppression of about two ($\RAA = 0.5$) is observed at low $\PT$
with $\RAA$ rising gently with increasing transverse momentum.  
The suppression is increasingly pronounced in the more-central collisions, 
as expected from the longer average path lengths traversed by hard-scattered partons as they lose energy via jet quenching.
In the 0--5\% bin, $\RAA$ reaches a minimum value of 0.13 around 6--7 GeVc.  At higher $\PT$, the value of $\RAA$ rises and levels off above 40 GeVc at a value of approximately 0.5.  

Measured $\RAA$ in the most central PbPb events (0--5 \%) is compared to a number of model predictions, both for the LHC design 
energy of $\sqrt{s_{_{\mathrm{NN}}}}$ = 5.5 TeV (PQM~\cite{pqm} and GLV~\cite{glv}) and for the actual 2010 collision energy of $\sqrt{s_{_{\mathrm{NN}}}}$ = 2.76 TeV (ASW, YaJEM, and an elastic scattering energy-loss model with parameterized escape probability~\cite{renk}).  While most models predict the generally rising behavior that is observed in the data at high $\PT$, the magnitude of the predicted slope varies greatly depending on the details of  the jet quenching implementation.

\section{Summary}
In this contribution, measurements of the charged particle transverse momentum spectra have been presented for 2.76 TeV PbPb collisions in bins of collision centrality. The results have been normalized to an interpolated 2.76 TeV reference $\PT$ spectrum to construct $\RAA$. A dramatic suppression (2--6 $\GeVc$) and a significant rise in the higher $\PT$ region with leveling-off around 40 $\GeVc$ of the charged particle spectrum have been observed for the most-central PbPb events. The new CMS measurement should help constrain the quenching parameters used in various models and further the understanding of the energy-loss mechanism.

\bibliography{auto_generated}   

\end{document}